\documentstyle[sprocl]{article}

\def\Journal#1#2#3#4{{#1} {\bf #2}, #3 (#4)}

\def\NCA{\em Nuovo Cimento}
\def\PRD{{\em Phys. Rev.} D}

\def\d{{\rm d}}
\def\gtrsim{{{\buildrel >\over\sim}}}
\def\lesssim{{{\buildrel <\over\sim}}}
\def\H{{\widehat H}}
\def\a{{\widehat a}}
\def\as{{{\widehat a}^*}}

\def\C{{\rm C}}
\def\R{{\rm R}}

\begin{document}

\title{THE PHYSICS OF NEGATIVE ENERGY DENSITIES}

\author{ADAM D. HELFER}

\address{Department of Mathematics, University of Missouri--Columbia,\\ 
Columbia, Missouri 65211, USA\\e-mail:  adam@godel.math.missouri.edu}

\maketitle\abstracts{I review some recent results showing that the
physics of negative energy densities, as predicted by relativistic
quantum field theories, is more complicated than has generally been
appreciated.  On the one hand, in external potentials where there is a
time--dependence, however slight, the Hamiltonians are unbounded below. 
On the other, there are limitations of quantum measurement in 
detecting or utilizing these negative energies.}

\section{Introduction}
It has been known for a long time that in some sense negative energy
densities are predicted by relativistic quantum field theory.  However,
it is only recently that we have begun to understand how often they
occur, or indeed the precise sense in which they occur.  
There are a number of surprises in the area, and it may not be too much
to say that we have been missing the essential physics of the
situation.

The main surprise is that arbitrarily negative energy densities occur
pervasively, even in tame, ``everyday'' situations.  Much of this paper
will outline how this occurs in a general quantum--field--theoretic
setting.  No one really knows, at present, how to resolve the
difficulties pervasive negative energy densities might cause.  I shall
outline, at the end, some suggestions of mine, and their possible
consequences, but this is very much an open area.

\section{Instances of Negative Energies, Densities and Fluxes}
Before discussing the recent general results, I recall some important
instances of negative energy density phenomena.

Casimir was probably the first to find a negative energy density in
quantum field theory.\cite{c}  
He showed that between two parallel perfect plane
conductors separated by a distance $l$ there was a renormalized energy 
\begin{equation}
  E=-\frac{\pi ^2\hbar c}{720  l^3}
\end{equation}
per unit area.  Although Casimir did not calculate the energy density,
one can show on invariance grounds, given the very simple nature of the
boundary conditions, that the expectation of the energy density is
constant, so
\begin{equation}
  \langle {\widehat T}_{tt}\rangle _{\rm ren}=
-\frac{\pi ^2\hbar c}{720  l^4}\, .
\end{equation}
I should mention at this juncture that, while the associated force of
attraction has been observed, this negative energy density has
{\em not.}  The force is essentially the component ${\widehat T}_{zz}$
of the stress--energy, whereas 
the energy density is ${\widehat T}_{tt}$.  There is a
relation between the two --- the long--time average of the force is
minus the gradient of the energy --- but we do not at present have a
measurement of the energy density operator, and it is hard to see
how such a measurement might be made with current technology. 
Detecting the effect of the negative energy by, say, a Cavendish
experiment would require plates of the order of a light--year on a
side.

A second instance of negative energy density was found by Hawking, at
the horizon of a black hole.\cite{h}  The argument here is indirect, but very
interesting because one sees the physical assumptions:  Hawking first
computed the famous blackbody spectrum of radiation from a black hole,
on the assumption that it had formed at some finite time in the past. 
Given that positive energy is being radiated to infinity, the energy
must come from somewhere, and the only source is the hole.  If the hole
is losing energy, and we may apply classical general relativity at the
horizon, then the Area Theorem (also due to Hawking) implies a negative
effective energy density as the horizon.\footnote{Recent computations
confirm this, and suggest that the energy density is negative (for
at least some observers) {\em everywhere} outside the
hole, becoming unambiguously positive only at infinity.\cite{v}}

A third instance of negative energy, more particularly of negative
energy fluxes, occurs in the radiation from moving mirrors, first
studied by Davies and Fulling.\cite{df,fd}  Here, in the case of one spatial
dimension, one has excellent analytic control over the model, so the
physical issues show up very clearly.  One finds in particular
correlations between negative and positive energy fluxes, and
restrictions on the negative fluxes.

A fourth instance of negative energies occurs with
``supercritical'' external potentials:  potentials strong enough to
engender pair creation.\cite{gmr}

Each of these examples is of interest in its own right, has stimulated a
great deal of work, and has led to a deeper understanding of quantum
field theory.  
On the other hand, it has not really been clear whether there is any
unified theory of negative energy densities, as
quantum--field--theoretic phenomena, although there has been some work
in this direction.  The seminal paper is by Epstein, Glaser and Jaffe.\cite{egj} 
It is a theorem in axiomatic field theory, and it asserts, essentially,
that if a theory is Poincar\'e--invariant, possesses a ground state, and
possesses an energy density operator $\widehat\rho$, then $\widehat\rho$
cannot be a non--negative operator.  In other words, the total energy
\begin{equation}
 \widehat{H} =\int \widehat\rho \,\d ^3 x
\end{equation}
is positive, but for {\em any} positive smooth compactly--supported
(that is, vanishing outside a bounded set) function $f(x,y,z)$, the
operator
\begin{equation}
 \widehat{H} (f)=\int \widehat\rho\, f\, \d ^3x
\end{equation}
has a spectrum including negative numbers.

While this result proves, for example, that the energy density operator
for the Klein--Gordon field in Minkowski space, if it exists, is not
positive, it bears no direct link to any of the instances of negative
energies above.  This is because none of those configurations is
Poincar\'e{}--invariant.  

These various occurences of negative energies are not the same
physically.  We shall have a better idea of this when we examine the
structures of the Hamiltonians, below.

\section{Some General Results}
In the past few years, some constructive results on the stress--energy have
become available.  These build on the work of many authors; I shall only
name Wald as one of the prime movers in this area, and direct the reader to his
recent book for other references.\cite{w}

Brunetti, Fredenhagen and K\"o{}hler showed that, for a linear Bose
field in curved space--time, the stress--energy operator was
well--defined (modulo a well--known c--number ambiguity) as an
operator--valued distribution.\cite{bfk}  What this means is that for any
tensor--valued test function $f^{ab}$, smooth {\em on space--time} and
with compact supports, the quantity
\begin{equation}
 \int {\widehat T}_{ab}\, f^{ab}\, \d \,{\rm vol}
\end{equation}
was a self--adjoint operator.  We shall see below that this result
is quite important.  At first blush, however, it seems
not quite one wants for the measurement of energy or the evolution
of the quantum fields.

One does not usually think of integrating the stress--energy
over a space--time volume; it has a more direct interpretation as
determining Hamiltonian operators
\begin{equation}
 {\widehat H}(\xi ,\Sigma )=\int {\widehat T}_{ab}\, \xi ^a\, \d\Sigma
^b\, .
\end{equation}
This operator is the generator of evolution along the vector field $\xi$ 
at the Cauchy surface $\Sigma$; it is a sort of weighted energy (or
momentum, or angular momentum, according to the character of $\xi ^a$)
operator at $\Sigma$.  For {\em these} operators, the results are
negative:\/\cite{adh1}

Suppose that $\xi ^a$ is not identically a Killing vector field at
$\Sigma$.  Then
\begin{enumerate}

\item[(a)] ${\widehat H}(\xi ,\Sigma )$ does not exist as a self--adjoint
operator.\footnote{A slightly weaker statement was proved in ref. 10; a full
proof is in ref. 11.} It does exist as a Hermitian form (that
is, its expectation values are well--defined on a dense set of states);
but

\item[(b)] ${\widehat H}(\xi ,\Sigma )$ is unbounded below; the set of states
on which its expectation value is $-\infty$ is dense in the physical
Hilbert space;

\item[(c)] ${\widehat H}(\xi ,\Sigma )$ does in a well--defined way generate
an evolution of the field algebra, but that evolution is not unitarily
implementable.

\end{enumerate}

All these phenomena occur whenever there is a departure, no matter how
slight, from $\xi ^a$ being the generator of a space--time symmetry.  In
particular, if a time--dependent gravitational field is present, no
matter how small, there are {\em no} timelike Killing vectors, and so
all of these phenomena are unavoidable.  They are all local, and can be
manifested as certain ultraviolet divergences.

In the next section, I will give a sketch of how such things can occur
in apparently innocuous situations.  For the moment, though, I want to
be quite explicit about what these results mean.  To some extent, they
relate to odd phenomena which have previously been encountered in
quantum field theory, but whose significances have not always been
thought out.

Property (a) means that the Hamiltonian cannot be interpreted as
an observable within conventional quantum theory.  (It is the
self--adjoint operators which have spectral representations;
Hermiticity, a weaker property for unbounded operators, does not
suffice.)  Operators which exist only as forms are known in other
contexts in quantum field theory, for example in current algebra.
However, it is a rather serious modification of the theory to say that
the Hamiltonian exists only weakly.

Property (b), that the expectations of the energy are unbounded
below, is perhaps the most serious.  In the ordinary course of physics,
one would suspect this to lead to instabilities and a breakdown of the
theory.  Indeed, something like this is known to happen in
``supercritical'' external potentials, where the na\"\i{}ve vacuum
becomes unstable against pair creation.  While there is more than a
passing connection between the present situation and supercritical
potentials, it should be emphasized that $\H (\xi ,\Sigma )$ is unbounded
below even in the presence of arbitrarily weak external fields.

A full discussion of the non--unitarity (property (c)) will be given
elsewhere; here I shall only make precise the sense in which it
occurs.

In quantum theory, as is well--known, there are two sorts of evolution:
that of the state vectors, and that of the
fields.  The state vectors are required to evolve by unitary motions
(except when reduction occurs).  However, the evolution of the fields is
determined by the field equations, and may or may not be unitarily
implementable (that is, achievable by conjugation with a unitary
operator on the Hilbert space).  It is often thought that these two
sorts of evolution are simply two sides of the same coin, and unitarity
in one sense implies it in the other, but this is incorrect.  

In fact, the physical occurence of non--unitarily implementable
evolutions has been known in some sense since the study of infrared
effects in quantum electrodynamics by Bloch and Nordsieck.\cite{bn}
The general picture that emerged from their analysis was that a
scattering involving only a finite exchange of energy could give rise
to a final state with infinitely many soft photons, and this infinity
could be ``big'' enough to take the final state out of the initial
Hilbert space.  In modern parlance, the evolution is not unitarily
implementable.  However, notice that this same argument would rule out
any {\em ultraviolet} non--unitarity.  What we have found is that this
last conclusion is wrong.  The argument for it is wrong because energy
is not a classical quantity, but an operator, and there are in general
complicated interference effects between positive-- and
negative--energy states.

\section{Structure of the Hamiltonians}
In this section, I shall examine the structure of the Hamiltonians of
linear bose theories to show how the phenomena discussed above appear. 
A rigorous treatment of this is technical, and the reader is
referred to the literature for that.\cite{adh1,adh4}  Here the goal is
conceptual.

Consider the usual energy operator
for a Klein--Gordon field in Minkowski space.  We may write this
operator in the form
\begin{equation}
{\widehat H} =\int {\widehat T}_{ab} t^a\d\Sigma ^b=
B^i{}_j \, {\hat a}^*_i \, {\hat a}^j\,
,
\end{equation}
where $\d\Sigma ^b =t^b\d x\d y\d z$, the 
${\hat a}^i$, ${\hat a}^*_i$ are annihilation and
creation operators, and $B^i{}_j$ are coefficients forming a
self--adjoint matrix.  (One can allow ``continuous matrices,'' too; this
distinction will be unimportant.  So in a momentum representation
$B^i{}_j\to B({\bf k},{\bf q})=\delta ({\bf k}-{\bf
q})(m^2+{\bf k}^2)^{1/2}$, as usual.)

Now imagine perturbing this situtation, but staying within the
category of linear field theories.  Such a perturbation might include:

\begin{itemize}

\item Allowing a small time-- and space--dependent gravitational field;

\item Allowing other sorts of time--dependent external potentials;

\item Measuring not exactly the energy, but some other weighted average
of the stress--energy operator.

\end{itemize}

\noindent The last might be interesting, for example, in problems in
quantum measurement theory.  Suppose we want to take into account the
fact that we can never perfectly synchronize clocks at different spatial
locations.  Then it seems we ought to allow perturbations of 
the vector field $t^a$ (and of the surface $\Sigma$).

If we allow such perturbations, then we find that the Hamiltonian will
have the general form
\begin{equation}
{\widehat H}=A_{ij}\a ^i\a ^j +B^i{}_j\as _i\a ^j +{\overline A}^{ij}
  \as _i\as _j +\mbox{c--number term}\, .
\end{equation}
Here $A_{ij}$ and $B^i{}_j$ are coefficients
(symmetric and self--adjoint, respectively).  The operators $\a$ and
$\as$ are still annihilation and creation operators, although in general
situations they might not refer to particles, but to other field modes. 
(The interpretation can only be determined by analysis of the particular
system and mode decomposition.)
The c--number term can be quite interesting --- it includes
Casimir--type effects --- but I shall have nothing more to say about it
here.  Thus the new features that we are concerned with are the
$A_{ij}$ terms.  The presence of these terms indicates that
evolution does not preserve the decomposition of field operators into
annihilation and creation parts.

In the case of perturbing a Klein--Gordon energy operator, one would
expect the coefficients $A_{ij}$ to be small, and of
$B^i{}_j$ to be close to those of the unperturbed case.  In all
of the cases considered here, this is true:  in a suitable sense, the
$A_{ij}$ are uniformly small, and the $B^i{}_j$
uniformly close to their unperturbed values.  However, just having the
coefficients uniformly small is not enough to rule out pathologies,
because in a field theory there are infinitely many
modes.\footnote{Another
difficulty is that the terms $A_{ij}\a ^i\a ^j$,
${\overline A}^{ij}\as _i\as _j$ are unbounded operators.}

There is a very simple example of this.  If $|0\rangle$ is the vacuum
state (that is, the state annihilated by all $\a$), then 
\begin{equation}
{\widehat H}|0\rangle ={\overline A}^{ij}\as _i\as _j|0\rangle\, .
\end{equation}
It is easy to see that this state is normalizable iff
$A_{ij}A^{ij}<\infty$.  However, this condition can
very well be violated, even if the $A_{ij}$ are uniformly
small.  It turns out that, if one calculates $A_{ij}$ in any of
the situations listed above, one does find $A_{ij}{\overline
A}^{ij}=\infty$, and the divergence is an ultraviolet one.  
This means that the vacuum cannot be in the
domain of $\widehat H$.
(By itself, this does not prove that the Hamiltonian cannot exist as a
self--adjoint operator.  One has to worry about whether there might be
some recondite domain of ``dressed'' states on which the Hamiltonian
turns out to be o.k.   However, it turns out that this does not occur.)

Now suppose for the moment we had a finite--dimensional system
instead of a field theory.  Then we would know that at the classical
level there was a canonical transformation taking the system to a
direct sum of harmonic oscillators.  This means that there is a
canonical basis of the classical phase space relative to which the
classical Hamiltonian vector field, which is
\begin{equation}
V=i\left[\begin{array}{cc} B&2\overline{A}\\ -2A&-B\end{array}\right]
\end{equation}
in a natural complex basis associated to the creation and annihilation
operators,
breaks down into a sum of $2\times 2$
blocks, each of the form $\left[\begin{array}{cc} 0&\omega\\ -\omega
&0\end{array}\right]$, with $\omega$ the angular frequency of the oscillator.  
If $V$ is the classical vector field, then, the ground--state energy in
terms of this new canonical basis is simply 
$(1/2)\sum _\omega \omega =(1/4){\rm tr}_\R\, |V|$,
where $|V|=\sqrt{-V^2}$.  Now, the ground--state energy in the
original basis is just $(1/2){\rm tr}_\C\, B =(1/4){\rm tr}_\R\, B$.  
Thus the renormalized
ground state energy is 
\begin{equation}
E_{\rm ground}={\rm tr}_\R\, \left( |V|-B\right) /4\, .
\end{equation}

In the case of fields, this formula is much more delicate to justify
rigorously, and indeed needs to be qualified in various ways.  Still, 
we might take the attitude that the cases in which this fails (and so
the field theory cannot be well--approximated by a sequence of
quantum mechanical theories) as in themselves pathological.  It turns
out that in many interesting
cases, the $A$'s are ``small'' and their commutators with $B$ are
negligible to first order, and then one finds simple approximations
$|V|\approx \sqrt{B^2-4{\overline A}A}$, and $E_{\rm ground}\approx
-(1/2){\rm tr}_\R B^{-1}\overline{A} A$.  This trace is ultraviolet
divergent in cases of interest.

The foregoing are sketches of how the lack of self--adjoint
implementability of $\H$ and unboundedness--below arise.
The proof of non--unitary implementability of evolution is a bit less
direct (since one has to make a statement about finite evolutions, but
it is too hard to integrate the equations of motion).

I will close this section with a few words about how these
negative--energy phenomena relate to others.  
As mentioned above,
Casimir--type effects are subsumed in the c--number term, and 
can be viewed as a kinematic rather than a dynamic effect.
In all known examples, they are locally finite.
The moving--mirror models are fields with (singular) time--dependent
potentials; they in effect have $A_{ij}\not= 0$.   
The case of Hawking
radiation is subtle, as the time--dependence is
is encoded in the
assumption that the black hole formed at some finite time;  
this model has $A_{ij}\not= 0$ as well.
For ``supercritical'' potentials, the modification of the theory is
severe enough that the present Hamiltonian structure breaks down
entirely.  This is clearest in the case of time--independent potentials.
For these, when one tries to construct the one--particle Hilbert space,
one finds modes with imaginary fundamental frequencies $\omega$. 
Formally speaking, in the time--independent case, the coefficients
$A_{ij}=0$ and $B^i{}_j$ acquires a non--self--adjoint part.  However,
this formal behavior is only a signal that the model has broken down;
one must look at the physics of the situation to understand what
actually happens.  (One must quantize in another representation of the
CCR, with a ``dressed'' vacuum.)

\section{Towards a Resolution of the Difficulties}
The most disturbing features of the negative energies uncovered above
are their unboundedness--below and their pervasiveness. Given these
features, why have we not already seen negative energy--density
phenomena?  Why do we not see ordinary particles absorb negative
energies and become tachyons?  Why do not small perturbations (which are
always present) send the field cascading through increasingly
negatively energetic states, while emitting positive--energy radiation? 
There must be some mechanism restricting negative energies, their
production, their duration, or their interaction with ordinary fields.

An important such restriction was discovered by Ford,
and elaborated by him and co--workers.\cite{fr}  Starting from
the physical observation that one does not measure energies (or energy
densities) at mathematical points, but always with some finite
averaging, he demonstrated that the stress--energy operator for the
Klein--Gordon field satisfies a {\em quantum inequality},
\begin{equation}
\langle\Psi |\int _{-\infty}^\infty {\widehat T}_{tt}(t,0,0,0)\,
   b(t)\, \d t|\Psi\rangle /\langle\Psi |\Psi\rangle \geq
  -3\hbar c/(32\pi ^2 (ct_0^4))\, ,
\end{equation}
where the sampling function $b(t)=(t_0/\pi)/(t^2+t_0^2)$ has
characteristic width $\sim t_0$ and area unity.\footnote{One can also
average over spatial directions; taking this into account would
strengthen all the arguments below.  However, the key thing turns out to
be temporal averaging.}  Thus one can have very negative energy
densities for short times, but, averaged over longer times, the energy
density must be more nearly zero.

These inequalities point up an important issue.  When we are
considering the quantum stress--energy, and especially whenever
negative energy--density phenomena are concerned, we should speak of
the energy in a {\em regime,} where this term includes not only a
spatial extent but a scale of temporal averaging.  It is generally
wrong to suppose that there is a well--defined energy density (even as
a quantum operator) independent of the temporal averaging scale.

I want now to combine the quantum inequality, which provides a
mathematical restriction on the stress--energy, with limitations of
quantum measurement theory.  Suppose an isolated device measures or
traps the energy of a quantum field in a regime.  If the device is to
trap a negative energy, then according to the quantum inequalities it
must turn on and off on a finite time scale, say $\sim t_0$, and so it
must contain a clock resolving times over order $\lesssim t_0$.  By
causality and the quantum inequality, the magnitude of any negative
energy detected or trapped is bounded by
\begin{equation}
|E_{\rm neg}|\leq \left[ 3\hbar c/(32\pi ^2 (ct_0^4))\right]\cdot 
  (4\pi /3)(ct_0)^4 =\hbar /(8\pi t_0)\, .
\end{equation}
Now there is an old argument,
going back at least to the Bohr--Einstein dialog,\cite{b} that a clock which
can resolve times of order $t_0$ must have a rest energy $\gtrsim
\hbar /t_0$.
This is about $25$ times as great as the negative energy detected!

This suggests a general principle, which I shall refer to by analogy
with the energy conditions of classical general relativity:  {\em
Operational Weak Energy Condition (OWEC):  The energy of a quantum
field in a regime, plus the energy of an isolated device in that regime
measuring or trapping the field's energy, must be non--negative.}

I would like to emphasize that while the argument for the OWEC is good,
it is not a rigorous proof.\footnote{Here are the caveats.  First,
the inequality $E_{\rm clock}\gtrsim\hbar /t_0$ is only known as
an order--of--magnitude relation.  Second, we do not at present have 
mathematically proved quantum inequalities in curved space--time, although 
there are excellent reasons for thinking that at a
local level there are similar results.  Third, it is not
possible to say anything rigorous about non--linear quantum fields. 
Finally, the OWEC could in principle be violated if there were a
sufficiently large number of elementary field species.}
Nevertheless, the form of the condition is so suggestive, and the
factor of $8\pi$ so in excess of unity, that the OWEC seems worthwhile
at least investigating as a hypothesis.

The OWEC would on its face rule out the conversion of ordinary particles
to tachyons by absorption of negative energies.  It is not yet known
whether the OWEC precludes other pathologies; one must investigate each
in turn.

Whether the OWEC rules out all negative--energy pathologies or not, it
is interesting to investigate as a candidate quantum analog of the
classical weak energy condition (WEC) in general
relativity.\footnote{The WEC asserts that the energy density meaured by
any oberver is non--negative.}  This condition is a foundation of the
most important results in classical relativity:  the Area Theorem, the
singularity theorems, and the positivity--of--energy theorems.

If the OWEC holds, and some version of Einstein's equation holds with
the stress--energy operator ${\widehat T}_{ab}$ as a source, then in a
negative--energy regime, one could never measure the geometry of
space--time sufficiently accurately by direct local means to establish
unambiguously the negativity of the energy density via Einstein's
equation.  In other words, {\em in a negative--energy regime, there are
quantum limitations on the measurement of space--time geometry.}  Given
the pervasiveness of negative energies, this suggests that ``quantum
space--time'' phenomena may be accessible at ordinary scales.  The
potential significance of this would be hard to overstate.

These ideas are discussed more extensively in refs. 15, 16.

\end{document}